\documentclass[a4paper,usenatbib]{mn2e}
\usepackage{amsmath}
\usepackage{graphicx}
\usepackage{url}

\usepackage[totalwidth=505pt,totalheight=680pt]{geometry}

\newlength{\wsmallfig}
\newlength{\wlargefig}
\newcommand{\sss}[1]{{\scriptscriptstyle{#1}}}
\newcommand{\order}[1]{\mathcal{O}\!\left(#1\right)}
\newcommand{\boldmathsymbol}[1]{{\ensuremath{\boldsymbol{#1}}}}

\newcommand{\ASPIC}{\texttt{ASPIC}}
\newcommand{\CAMB}{\texttt{CAMB}}
\newcommand{\COSMOMC}{\texttt{COSMOMC}}
\newcommand{\CAMSPEC}{\texttt{CamSpec}}
\newcommand{\MULTINEST}{\texttt{MULTINEST}}
\newcommand{\CLIK}{\texttt{CLIK}}
\newcommand{\FIELDINF}{\texttt{FieldInf}}
\newcommand{\SKYNET}{\texttt{SKYNET}}

\newcommand{\LCDM}{\Lambda\mathrm{CDM}}

\newcommand{\ee}{e}
\newcommand{\efold}{$\ee$-fold}
\newcommand{\efolds}{$\ee$-folds}

\newcommand{\uS}{\mathrm{S}}
\newcommand{\ud}{\mathrm{d}}

\newcommand{\uinf}{\mathrm{inf}}

\newcommand{\uend}{\mathrm{end}}
\newcommand{\urad}{\mathrm{rad}}
\newcommand{\ureh}{\mathrm{reh}}
\newcommand{\unuc}{\mathrm{nuc}}

\newcommand{\uc}{\mathrm{c}}
\newcommand{\up}{\mathrm{p}}
\newcommand{\ub}{\mathrm{b}}
\newcommand{\ut}{\mathrm{t}}
\newcommand{\uk}{\mathrm{k}}
\newcommand{\udm}{\mathrm{dm}}
\newcommand{\ueff}{\mathrm{eff}}

\newcommand{\uSZ}{\mathrm{SZ}}
\newcommand{\usssMC}{\sss{\mathrm{MC}}}
\newcommand{\usssPS}{\sss{\mathrm{PS}}}
\newcommand{\usssCIB}{\sss{\mathrm{CIB}}}
\newcommand{\utSZ}{\ut\sss{\uSZ}}
\newcommand{\ukSZ}{\uk\sss{\uSZ}}

\newcommand{\rdof}{\mathcal{Q}}
\newcommand{\rdofreh}{\rdof_\ureh}
\newcommand{\gzero}{g_\sss{0}}
\newcommand{\greh}{g_\ureh}
\newcommand{\gs}{q}
\newcommand{\gszero}{\gs_\sss{0}}
\newcommand{\gsreh}{\gs_\ureh}
\newcommand{\Mpl}{M_{_\mathrm{P}}}

\newcommand{\wrehbar}{\overline{w}_{\ureh}}

\newcommand{\nS}{n_{\sss{\uS}}}

\newcommand{\kstar}{k_*}
\newcommand{\etastar}{\eta_*}
\newcommand{\eps}[1]{\epsilon_{#1}}
\newcommand{\Pstar}{P_*}
\newcommand{\Hstar}{H_*}
\newcommand{\Hzero}{H_{\sss{0}}}
\newcommand{\epsstar}[1]{\eps{#1*}}
\newcommand{\epsend}{\eps{1\uend}}

\newcommand{\calL}{\mathcal{L}}
\newcommand{\calP}{\mathcal{P}}
\newcommand{\calPh}{\calP_h}
\newcommand{\calPz}{\calP_\zeta}

\newcommand{\MeV}{\mathrm{MeV}}
\newcommand{\GHz}{\mathrm{GHz}}
\newcommand{\Mpc}{\mathrm{Mpc}}

\newcommand{\post}[2]{P\negthinspace\left(#1|#2\right)}

\newcommand{\calLeff}{\calL_\ueff}
\newcommand{\likeff}[2]{\calLeff\negthinspace\left(#1|#2\right)}
\newcommand{\likeffb}[2]{\calLeff\negthinspace\left[#1|#2\right]}

\newcommand{\thetainf}{\theta_{\uinf}}
\newcommand{\thetaMC}{\theta_{\usssMC}}

\newcommand{\bthetainf}{\boldmathsymbol{\theta}_{\uinf}}
\newcommand{\bthetareh}{\boldmathsymbol{\theta}_{\ureh}}
\newcommand{\bthetaprim}{\boldmathsymbol{\theta}{_{\up}}}
\newcommand{\bthetacosmo}{\boldmathsymbol{\theta}_{\uc}}
\newcommand{\bdata}{\boldmathsymbol{D}}
\newcommand{\bk}{\boldmathsymbol{k}}

\newcommand{\OmegaL}{\Omega_\Lambda}
\newcommand{\OmegaCDM}{\Omega_\udm}
\newcommand{\OmegaB}{\Omega_\ub}

\newcommand{\APSa}{A^{\usssPS}_{100}}
\newcommand{\APSb}{A^{\usssPS}_{143}}
\newcommand{\APSc}{A^{\usssPS}_{217}}
\newcommand{\rPSbc}{r^{\usssPS}_{143\times217}}
\newcommand{\ACIBb}{A^{\usssCIB}_{143}}
\newcommand{\ACIBc}{A^{\usssCIB}_{217}}
\newcommand{\rCIBbc}{r^{\usssCIB}_{143\times217}}
\newcommand{\gamCIB}{\gamma^{\usssCIB}}
\newcommand{\AtSZ}{A_{\utSZ}}
\newcommand{\AkSZ}{A_{\ukSZ}}
\newcommand{\xitSZCIB}{\xi^{\utSZ\times\usssCIB}}

\newcommand{\ca}{c_{100}}
\newcommand{\cc}{c_{217}}
\newcommand{\betaoo}{\beta_1^1}

\newcommand{\areh}{a_\ureh}
\newcommand{\aend}{a_\uend}
\newcommand{\xend}{x_\uend}
\newcommand{\xstar}{x_*}
\newcommand{\Nend}{N_\uend}
\newcommand{\Nstar}{N_*}
\newcommand{\Nzero}{N_0}
\newcommand{\Rrad}{R_\urad}
\newcommand{\Rreh}{R}
\newcommand{\Vend}{V_\uend}
\newcommand{\Vstar}{V_*}
\newcommand{\vstar}{v_*}
\newcommand{\zend}{z_\uend}

\newcommand{\rhotildegamma}{\tilde{\rho}_\gamma}
\newcommand{\rhogamma}{\rho_\gamma}

\newcommand{\rhoreh}{\rho_\ureh}
\newcommand{\rhoend}{\rho_\uend}

\newcommand{\rhonuc}{\rho_\unuc}

\title[Fast Bayesian inference for slow-roll inflation]{Fast Bayesian inference for slow-roll inflation}
\author[Christophe Ringeval]{Christophe
  Ringeval\thanks{christophe.ringeval@uclouvain.be} \\
Centre for Cosmology, Particle Physics and Phenomenology \\
Institute of Mathematics and Physics, Louvain University \\
2 chemin du cyclotron, 1348 Louvain-la-Neuve, Belgium}

\begin{document}

\makeatletter
\def\mniiiauthor#1#2#3{%
  \@ifundefined{mniiiauth@#1}
    {\global\expandafter\let\csname mniiiauth@#1\endcsname\null #2}
    {#3}}
\makeatother

\date{\today}
\pagerange{3253--3261}
\pubyear{2014}
\volume{439}

\maketitle

\label{firstpage}

\begin{abstract}
We present and discuss a new approach increasing by orders of
magnitude the speed of performing Bayesian inference and parameter
estimation within the framework of slow-roll inflation. The method
relies on the determination of an effective likelihood for inflation
which is a function of the primordial amplitude of the scalar
perturbations complemented with the necessary number of the so-called
Hubble flow functions to reach the desired accuracy. Starting from any
cosmological data set, the effective likelihood is obtained by
marginalisation over the standard cosmological parameters, here viewed
as ``nuisance'' from the early Universe point of view. As being
low-dimensional, basic machine-learning algorithms can be trained to
accurately reproduce its multidimensional shape and then be used as a
proxy to perform fast Bayesian inference on the inflationary
models. The robustness and accuracy of the method are illustrated
using the Planck Cosmic Microwave Background (CMB) data to perform
primordial parameter estimation for the large field models of
inflation. In particular, marginalised over all possible reheating
history, we find the power index of the potential to verify $p<2.3$ at
$95\%$ of confidence.

\end{abstract}


\section{Introduction}
The recent release of the Planck CMB data~\citep{Ade:2013ydc,
  Ade:2013xla} and of the small scales CMB
experiments~\citep{Hou:2012xq, Dunkley:2013vu, Sievers:2013ica} have
secured some generic predictions of inflationary cosmology: a very
small spatial curvature today and an almost scale invariant primordial
power spectrum for the cosmological perturbations. \cite{Ade:2013zuv}
reports a spectral index $\nS=0.9603 \pm 0.0073$ using Planck
temperature data complemented with WMAP
polarization~\citep{Bennett:2012zja, Hinshaw:2012aka}. Testing cosmic
inflation is both a theoretical and experimental
challenge. Originally, inflation was introduced to solve the problems
of the standard Friedmann-Lema\^{\i}tre model of cosmology and this
requires a period of accelerated expansion to take place in the early
Universe~\citep{Starobinsky:1980te, Guth:1980zm, Linde:1981mu,
  Albrecht:1982wi}. As such, inflation used to be referred to as a
paradigm being difficult to test in itself as it could be implemented
in various ways. However, getting the right spectrum for the
cosmological perturbations requires the accelerated expansion to be
complemented with some self-gravitating scalar degree of freedom
experiencing quantum fluctuations~\citep{Starobinsky:1979ty,
  Mukhanov:1981xt,Mukhanov:1982nu, Bardeen:1983qw, Mukhanov:1985rz,
  Goncharov:1987ir, Mukhanov:1988jd, Mukhanov:1990me}. In its simplest
implementation, such an evolution can be obtained with a single scalar
field $\phi$ slowly rolling down a flat enough potential. These
slow-roll models come with additional generic predictions, such as a
small amount of non-Gaussianities and adiabatic initial conditions for
the cosmological perturbations, both being in agreement with current
cosmological data~\citep{Ade:2013ydc, Ade:2013xla}. As discussed by
\cite{Martin:2013tda}, the slow-roll class is already a populated
landscape of well-motivated theoretical scenarios which make definite
observable predictions. As such, testing single field inflation is
challenging not because the models are too generic but because there
are a very large number of different scenarios.

In this context, there are various ways to confront the slow-roll
models with cosmological data. The basic approach consists in
comparing crude predictions for the spectral index $\nS$ and the
tensor-to-scalar ratio $r$ to existing bounds derived from a data
analysis based on power law primordial power spectra. There are
various issues with such an approach. First, the observable effects
coming from the duration of the reheating era are omitted whereas,
since the WMAP data, those are known to be relevant for many
models~\citep{Martin:2006rs, Martin:2010kz}. Secondly, predicting the
spectral index, for a given inflationary model, usually relies on a
first order expansion in a small parameter $\epsilon$, proportional to
the field velocity squared (in {\efold} time), from which one can show
that $\nS - 1 = \order{\epsilon}$. For a value of $\nS \simeq 0.96$,
one has $\order{\epsilon^2} \simeq 2 \times 10^{-3}$ and one can
question the relevance of the second order terms in view of the very
small error bars on the current measurements on $\nS$ (see
above). Finally, let us stress that the true power spectra steaming
from a given inflationary model do not generically have a power law
shape. As such, one could also question the relevance of using
$(\nS,r)$ as a proxy to constrain the inflationary power
spectra. Those questions used to be of negligible interest in a recent
past as their effects were considered much smaller than the
observational uncertainties. But as just argued, that is no longer the
case with the Planck satellite results.

A way to alleviate these uncertainties is to numerically integrate the
cosmological perturbations mode by mode during
inflation~\citep{Salopek:1988qh, Adams:2001vc, Makarov:2005uh,
  Ringeval:2007am}. In that situation, for a given theoretical model,
one obtains the exact primordial power spectra for both the tensor and
scalar perturbations which depend on both the inflationary parameters
$\{\bthetainf\}$ and the reheating parameters $\{\bthetareh\}$. For
instance, considering a free massive scalar field of mass $m$ to be
the inflaton, the equations of motion for the perturbations in Fourier
space involve both $\thetainf=m$ and the wavenumber $k$. The reheating
parameters $\{\bthetareh\}$ indirectly appear through the mapping
between the physical wavenumbers observed today $k/a_0$ and the actual
value of $k/a$ during inflation. As a result, for any model of
inflation, the power spectra have to depend on both $\{\bthetainf\}$
and $\{\bthetareh\}$. The method being exact, it is readily applicable
to multifield inflation or more exotic models. Performing CMB data
analysis from an exact numerical integration has been implemented for
the first time by~\cite{Ringeval:2005yn, Martin:2006rs} and gives
marginalised posterior distributions directly on to the fundamental
parameters $\{\bthetainf,\bthetareh\}$ we are interested in. Moreover,
the effects coming from the reheating being necessarily taken into
account, this has been used to obtain the first CMB constraints on the
reheating history in~\cite{Martin:2010kz}. Although now routinely
used, see for instance~\cite{Mortonson:2010er,
  Ade:2013uln}, numerically integrating the inflationary spectra is
computationally demanding. As a result, performing parameter
estimation and Bayesian inference with this technique remains limited
to a small number of models only~\citep{Martin:2010hh,
  Easther:2011yq}.

In between, a precise way to perform CMB data analysis within
slow-roll inflation has been discussed by \cite{Leach:2002dw}. It
relies on the analytic expression of the primordial power spectra
that can be consistently derived within the so-called slow-roll
approximation~\citep{Mukhanov:1985rz, Stewart:1993bc, Martin:1999wa,
  Schwarz:2001vv, Leach:2002ar, Schwarz:2004tz}. Defining the Hubble
flow functions~\citep{Hoffman:2000ue, Schwarz:2001vv} (also named
slow-roll parameters) by
\begin{equation}
\eps{i+1} \equiv \dfrac{\ud \ln |\eps{i}|}{\ud N}, \qquad
\eps{1} \equiv -\dfrac{\ud \ln H}{\ud N}\,,
\label{eq:hubbleflow}
\end{equation}
with $H=\dot{a}/a$, $N\equiv \ln a$, $a$ being the
Friedmann-Lema\^{\i}tre-Robertson-Walker (FLRW) scale factor, one can
check that the expansion of the Universe is accelerated when
$\eps{1}<1$ ($\ddot{a}>0$). If the dynamics of the Universe is
dominated by a scalar field, one can moreover show that $\eps{1} =
(1/2) (\ud \phi/\ud N)^2$ (in Planck units), i.e. $\eps{1}$ is
directly proportional to the field velocity squared. The slow-roll
approximation assumes that all the $\eps{i}\ll 1$ and are, at most, of
the same order $\order{\epsilon}$. In that situation, the primordial
power spectra can be derived analytically, at a given order in
$\order{\epsilon}$. These calculations are non-trivial as they consist
in solving the equations of motion for both the scalar and tensor
perturbations during inflation. Currently, they have been completely
performed up to second order~\citep{Stewart:1993bc,
  Gong:2001he,Leach:2002ar, Martin:2002vn, Habib:2002yi, Habib:2004kc,
  Casadio:2005em, Casadio:2005xv, Lorenz:2008et, Martin:2013uma,
  Jimenez:2013xwa} and one obtains
\begin{equation}
\begin{aligned}
& \calPh = \dfrac{2 \Hstar^2}{\pi^2 \Mpl^2} \left\{ 1 -
  2(1+C)\epsstar{1} +\left(\dfrac{\pi^2}{2}-3 + 2C +2C^2 \right)
  \epsstar{1}^2 \right.\\ & + \left. \left(\dfrac{\pi^2}{12} -2 -2C
  -C^2 \right) \epsstar{1} \epsstar{2} + \left[ -2 \epsstar{1} +
    (2+4C) \epsstar{1}^2 \right. \right. \\ &- \left. \left. 2(1+C)
    \epsstar{1} \epsstar{2} \right] \ln\left(\dfrac{k}{\kstar}\right)
  + \left(2 \epsstar{1}^2 - \epsstar{1}\epsstar{2} \right)
  \ln^2\left(\dfrac{k}{\kstar}\right) \right\},
\end{aligned}
\label{eq:powerh}
\end{equation}
for the tensor modes and
\begin{equation}
\begin{aligned}
& \calPz  = \dfrac{\Hstar^2}{8 \pi^2 \Mpl^2 \epsstar{1}} \bigg\{ 1
-2(1+C)\epsstar{1} - C \epsstar{2} \\ & + \left(\dfrac{\pi^2}{2} -3 +
2C + 2C^2 \right) \epsstar{1}^2 +\left(\dfrac{\pi^2}{24} -
\dfrac{C^2}{2} \right) \epsstar{2} \epsstar{3} \\ & + \left( \dfrac{7
  \pi^2}{12} -6 -C+C^2\right) \epsstar{1} \epsstar{2} +
\left(\dfrac{\pi^2}{8} -1 +\dfrac{C^2}{2} \right) \epsstar{2}^2 \\ & +
\bigg[-2 \epsstar{1} - \epsstar{2} +(2+4C) \epsstar{1}^2 + (-1 + 2C)
  \epsstar{1} \epsstar{2} \\&+ C \epsstar{2}^2 - C \epsstar{2}
  \epsstar{3} \bigg] \ln\left(\dfrac{k}{\kstar}\right) + \bigg(2
  \epsstar{1}^2 + \epsstar{1} \epsstar{2} + \dfrac{1}{2} \epsstar{2}^2
  \\ & - \dfrac{1}{2} \epsstar{2} \epsstar{3} \bigg) \ln^2
\left(\dfrac{k}{\kstar}\right) \bigg\},
\end{aligned}
\label{eq:powerz}
\end{equation}
for the comoving curvature perturbation. In these expressions, all
quantities with a ``$*$'' are functions evaluated at the conformal time
$\etastar$ defined by
\begin{equation}
\kstar \etastar = -1,
\end{equation}
i.e., the time at which a pivot mode of astrophysical interest today,
for instance $\kstar=0.05\,\Mpc^{-1}$, crossed the Hubble radius
during inflation. The quantity $C=\gamma+\ln(2)-2 \simeq -0.72964$, where
$\gamma$ is the Euler constant, stems from the integration of the
equations of motion. These expressions are not exactly of a power law
shape and show that performing inflationary data analysis using
$\calPz \propto (k/\kstar)^{\nS-1}$ will necessarily introduce some
bias of $\order{\epsilon^2}$.

Taking equations~\eqref{eq:powerh} and \eqref{eq:powerz} as an input
for cosmological data analysis consists in assuming that slow-roll
inflation can be accurately described by the set of parameters
$(\Hstar, \epsstar{1}, \epsstar{2}, \epsstar{3},\dots)$. This is a
fair assumption precisely because the Hubble flow hierarchy is
constructed to do so, and any desired accuracy can be reached by
including higher order terms. As an outcome, one obtains the marginalised
probability distributions on $\Hstar$ and the slow-roll parameters
$\epsstar{i}$~\citep{Leach:2002dw, Martin:2006rs, Lorenz:2008je,
  Finelli:2009bs, Kuroyanagi:2013ns}. In fact, it turns out to be more
convenient from a data analysis point of view to trade $\Hstar$ for
the quantity $\Pstar \equiv \calPz(\kstar)$, i.e. the amplitude of the
primordial anisotropies at the pivot scale. From
equation~\eqref{eq:powerz}, we see that $\Hstar$ is indeed uniquely
determined given $\{\Pstar,\epsstar{i}\}$.

Compared to an exact numerical integration, this is not really what we
would like to obtain as the parameters we are interested in, for a given
model of inflation, are $\{\bthetainf,\bthetareh\}$. However, once the
potential is specified, the slow-roll approximation allows us to map,
order by order, all the Hubble flow functions $\eps{i}$ to the
successive derivatives of the
potential~\citep{Liddle:1994dx}. Therefore, within slow-roll, the
functions $\epsstar{i}(\bthetainf,\bthetareh)$ can be uniquely
determined, as it is for the amplitude
$\Pstar(\bthetainf,\bthetareh)$. In fact, their analytic expressions
for most of the single field models proposed so far can be found
in~\cite{Martin:2013tda} and the associated public code
$\ASPIC$\footnote{\url{http://cp3.irmp.ucl.ac.be/~ringeval/aspic.html}}.

In this paper, we show that using equations~\eqref{eq:powerh} and
\eqref{eq:powerz} to extract an effective likelihood in the slow-roll
variables space $(\Pstar,\epsstar{i})$, complemented by the knowledge
of the functionals $\epsstar{i}(\bthetainf,\bthetareh)$ and
$\Pstar(\bthetainf,\bthetareh)$, is enough to accurately constrain the
inflationary parameters $\bthetainf$ and $\bthetareh$ of any
slow-rolling inflationary model. Such an approach has the advantage of
requiring only one complete analysis of the cosmological data sets
under scrutiny, precisely to evaluate the effective likelihood, as
opposed to one per model for an exact numerical integration. Moreover,
the slow-roll approximation allows us to shortcut any mode integration
and the determination of the actual values of
$\epsstar{i}(\bthetainf,\bthetareh)$ for any model consists in solving
an algebraic equation for the reheating parameters. In
section~\ref{sec:method}, we show how to practically implement this
method using the publicly available codes
$\COSMOMC$~\citep{Lewis:1999bs, Lewis:2002ah} and
$\MULTINEST$~\citep{Feroz:2007kg, Feroz:2008xx, Trotta:2008bp,
  Feroz:2013hea}, together with a basic machine-learning
algorithm. Section~\ref{sec:practical} assesses its accuracy using the
Planck 2013 data~\citep{Ade:2013ktc} for the large field models of
inflation. In particular, we show that there are no significant
differences between the posterior distributions obtained from our
approach and the posteriors steaming from a mode by mode exact
integration of the inflationary perturbations pipelined with the exact
likelihood provided by the Planck
collaboration~\citep{Planck:2013kta}. Finally, we conclude in
section~\ref{sec:conclusion} and discuss how this approach could be
generalized to any models of inflation.

\section{Method}
\label{sec:method}

The first step consists in determining an effective likelihood in the
inflationary parameter space. In the framework of Bayesian statistics,
making inference on the parameters $\{\bthetainf\}$ and
$\{\bthetareh\}$ is done by marginalisation over all the others, see
for instance~\cite{Trotta:2008qt}. For CMB data, the marginalisation
is therefore performed over the standard cosmological and
astrophysical parameters, which are viewed as ``nuisance'' from the
early Universe point of view~\citep{Bridle:2001zv}. For instance, for
a flat $\Lambda$ Cold Dark Matter ($\Lambda$CDM) Universe, the
standard cosmological parameters are $\{\bthetacosmo\}=\{\OmegaB h^2,
\OmegaCDM h^2, \tau, \thetaMC\}$ where $\OmegaB$ is the density
parameter of baryons, $\OmegaCDM$ of cold dark matter, $\tau$ is the
Thomson scattering optical depth, $\thetaMC$ encodes the angular size
of the sound horizon at last scattering~\citep{Lewis:2002ah} and $h$
the reduced Hubble constant today. The marginalised posterior
probability distribution for the primordial parameters
$\{\bthetaprim\} \equiv\{\bthetainf,\bthetareh\}$, given some data set
$\bdata$ and some prior information $I$, reads
\begin{equation}
\post{\bthetaprim}{\bdata,I} = \int
  \post{\bthetaprim,\bthetacosmo,\bvarepsilon}{D,I} \ud
  \bvarepsilon \ud \bthetacosmo,
\label{eq:marge}
\end{equation}
where we have made explicit the standard cosmological parameters
$\{\bthetacosmo\}$ and have introduced a set of auxiliary parameters
$\{\bvarepsilon\}=\{\varepsilon_0,\varepsilon_1,\dots\}$. These
auxiliary parameters allow us to extend the inference problem to the
slow-roll parameter space, in a way similar to the introduction of
hyperparameters~\citep{Hobson:2002zf}. For instance, we can choose to
identify $\varepsilon_0$ to the scalar amplitude $\Pstar$ and the
$\varepsilon_i$ to the Hubble flow hierarchy $\epsstar{i}$. As
discussed in the introduction, within a particular model of slow-roll
inflation, the primordial amplitude and the slow-roll parameters are
deterministic variables, i.e.
\begin{equation}
\post{\bvarepsilon}{\bthetaprim,I} = \delta\left[\varepsilon_0 -
\Pstar(\bthetaprim)\right] \prod_{i\ge1} \delta\left[\varepsilon_i -
  \epsstar{i}(\bthetaprim) \right].
\label{eq:delta}
\end{equation}
From the Bayes' theorem, the joint probability distribution in the
right hand side of equation~\eqref{eq:marge} can be expanded in
\begin{equation}
\post{\bthetaprim,\bthetacosmo,\bvarepsilon}{\bdata,I} = \dfrac{
  \post{\bthetaprim,\bthetacosmo,\bvarepsilon}{I}
  \post{\bdata}{\bthetaprim,\bthetacosmo,\bvarepsilon,I}}{\post{\bdata}{I}}\,.
\end{equation}
Moreover, using the product rule and equation~\eqref{eq:delta}, one
has
\begin{equation}
\begin{aligned}
& \post{\bthetaprim,\bthetacosmo,\bvarepsilon}{I}  =
\post{\bthetaprim,\bthetacosmo}{I}
\post{\bvarepsilon}{\bthetaprim,\bthetacosmo,I} \\ & =
\post{\bthetaprim}{I}\post{\bthetacosmo}{I} \delta\left[\varepsilon_0 -
\Pstar(\bthetaprim)\right]\prod_i \delta\left[\varepsilon_i -
  \epsstar{i}(\bthetaprim) \right],
\end{aligned}
\end{equation}
since $\{\bthetaprim\}$ and $\{\bthetacosmo\}$ are independent
parameter sets. Using these expressions, equation~\eqref{eq:marge}
simplifies to
\begin{equation}
\post{\bthetaprim}{\bdata,I} = \dfrac{
  \likeffb{\bdata}{\Pstar(\bthetaprim),\epsstar{i}(\bthetaprim),I}
  \post{\bthetaprim}{I} }{\post{\bdata}{I}}\,,
\label{eq:margeff}
\end{equation}
where the marginalised effective likelihood is defined by
\begin{equation}
\begin{aligned}
&\likeff{\bdata}{\Pstar,\epsstar{i},I} =  \int
  \post{\bdata}{\bthetacosmo, \Pstar, \epsstar{i},I}
  \post{\bthetacosmo}{I} \ud \bthetacosmo.
\end{aligned}
\label{eq:likeff}
\end{equation}
Notice that the primordial parameters $\{\bthetaprim\}$ do no longer
appear explicitly in the likelihood because, within the slow-roll
approximation, the primordial power spectra are given by
equations~\eqref{eq:powerh} and \eqref{eq:powerz} such that the
parameters that may affect the likelihood are $\{\Pstar,\epsstar{i}\}$
only. In some way, we are using the slow-roll functional shape of the
primordial spectra to compress all of the available information into a
minimal number of parameters.

In practice, one should first evaluate $\calLeff$ by marginalisation
of the full likelihood using equation~\eqref{eq:likeff}. This requires
a complete data analysis including the standard cosmological and
astrophysical parameters, and can be computationally demanding, but
this has to be done once and for all. Once $\calLeff$ is determined,
any slow-roll inflationary models can be dealt with
equation~\eqref{eq:margeff}. The dimension of the parameter space
being reduced, depending on how fast one can evaluate $\calLeff$, the
speed of performing Bayesian inference and primordial parameter
estimation can be significantly increased.

In the next section, we put these considerations into practice and use
the Planck 2013 CMB data to determine $\calLeff$. Then, we apply our
method to some typical inflationary models and compare the results to
the ones coming from an exact numerical integration.

\section{Practical implementation}
\label{sec:practical}

In order to determine $\calLeff$, we have first performed a complete
CMB data analysis starting from the primordial power spectra given in
equations~\eqref{eq:powerh} and \eqref{eq:powerz}, i.e. expanded at
second order in the Hubble flow functions, that we now describe.

\subsection{Slow-roll analysis of the Planck CMB data}
\label{sec:srplanck}
The full likelihood $\post{\bdata}{\Pstar,\epsstar{i},\bthetacosmo}$
is computed using the publicly available $\CLIK$ code provided by the
Planck collaboration~\citep{Ade:2013ktc, Planck:2013kta}. The Planck
likelihood takes as an input the theoretical angular power spectrum of
the CMB anisotropies, $C_\ell$, for the polarization and the
temperature, together with additional sets of astrophysical and
observational parameters required to fit the foregrounds and various
instrumental nuisances. In order to determine the $C_\ell$ given our
cosmological and slow-roll parameters, we have used a modified version
of the $\CAMB$ code~\citep{Lewis:1999bs} to integrate the cosmological
perturbations starting from the initial conditions given by our power
spectra~\eqref{eq:powerh} and \eqref{eq:powerz}. More specifically,
the underlying likelihood is the so-called $\CAMSPEC$ likelihood,
described in~\cite{Planck:2013kta}, and the set of all cosmological,
astrophysical and nuisance parameters is of dimension eighteen:
\begin{equation}
\begin{aligned}
\{\bthetacosmo\} = & \left\{\OmegaB h^2, \OmegaCDM h^2, \tau,
100\thetaMC, \right. \\ & \left. \APSa, \APSb, \APSc, \rPSbc, \ACIBb,
\ACIBc, \rCIBbc, \right. \\ & \left. \gamCIB, \AtSZ, \AkSZ, \xitSZCIB,
\ca, \cc, \betaoo \right\}.
\end{aligned}
\label{eq:camparams}
\end{equation}
The first four parameters have been described before and are the
standard $\LCDM$ parameters. The next four respectively measure the
power contribution at $\ell=3000$ of unresolved point sources at
$100\,\GHz$, at $143\,\GHz$, at $217\,\GHz$ and their cross
correlation. The next three are their equivalent for the Cosmic
Infrared Background (CIB), and $\gamCIB$ stands for the spectral index
of the CIB angular power spectrum. The next three parameters measure the
unresolved Sunyaev-Zel'dovich contribution, thermal and kinetic and
its correlation with the CIB. Finally, the last three parameters
ensure marginalisation over calibration and beam uncertainties. More
details on the modelling of all these signals can be
found~\cite{Planck:2013kta}. In addition to the $\bthetacosmo$, our
data analysis adds the four-dimensional set of slow-roll parameters
$\{\Pstar,\epsstar{1},\epsstar{2},\epsstar{3}\}$.

Sampling the full parameter space has been performed with Markov chain
Monte Carlo (MCMC) methods by using the publicly available $\COSMOMC$
code~\citep{Lewis:2002ah}, and under the same hypothesis described
in~\cite{Ade:2013zuv}. In particular, our data sets are the Planck
temperature measurements complemented by the WMAP
polarization~\citep{Bennett:2012zja}, while all priors for the
cosmological parameters $\{\bthetacosmo\}$ have been kept identical to
those used by the Planck collaboration, see Table~$4$
in~\cite{Ade:2013zuv}. Concerning the slow-roll parameters, we have
chosen a Jeffreys' prior for the amplitude and for the first slow-roll
parameter: $\ln(10^{10}\Pstar)\in[2.7,3.4]$ and
$\log(\epsstar{1})\in[-5,0]$. Indeed, the order of magnitude for those
is a priori unknown. For the second and third slow-roll parameter, we
have chosen a flat prior in $[-0.2,0.2]$, motivated by the fact that
both should be less than one.
\begin{figure}
\begin{center}
\includegraphics[width=\wsmallfig]{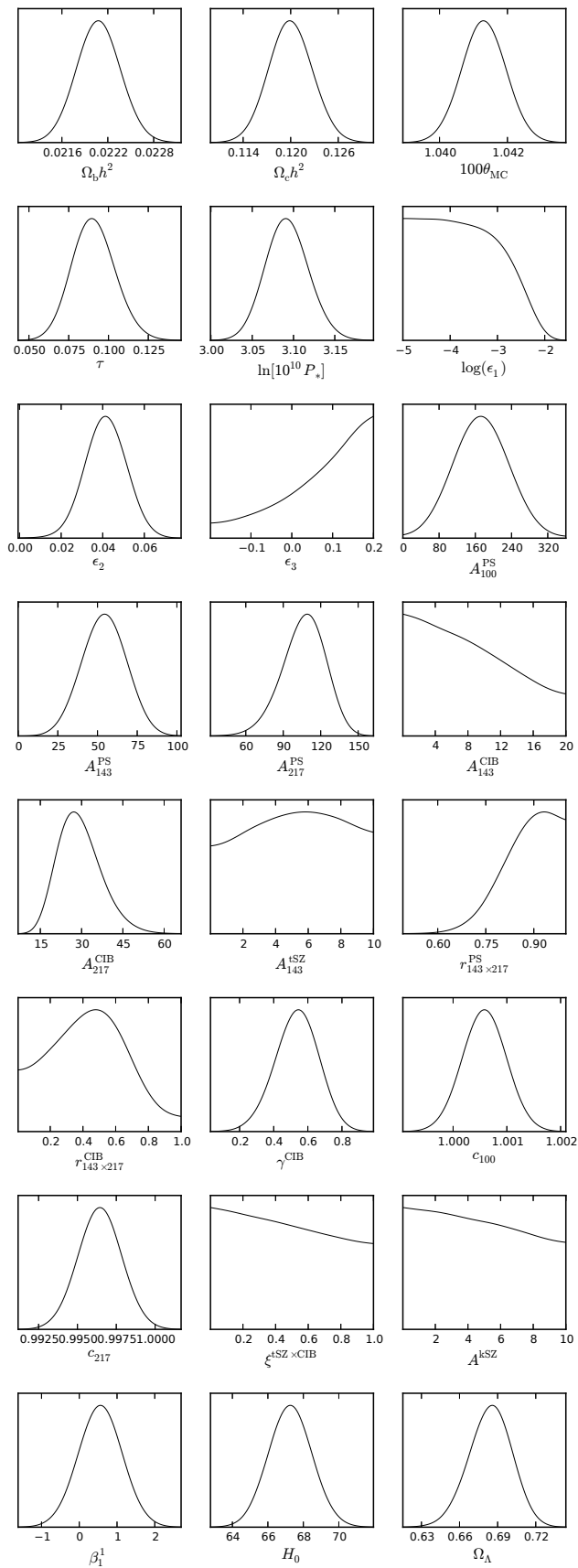}
\caption{Marginalised posterior probability distribution for all the
  parameters associated with our CMB slow-roll analysis of the Planck
  2013 data.}
\label{fig:cmb}
\end{center}
\end{figure}
The MCMC exploration has been stopped according to the R-statistics
convergence criteria implemented in $\COSMOMC$~\citep{Lewis:2002ah} and
this merely requires the variance between mean values computed within
each chain to be small enough. In our case, we have required it to be
smaller than $0.001$ and this amounts to keep a total number of
samples around $2\times 10^6$. The marginalised posterior probability
distributions for all the parameters have been plotted in
figure~\ref{fig:cmb}. In the space of $\{\bthetacosmo\}$, we recover
the same results as the Planck collaboration~\citep{Ade:2013zuv}. For
completeness, we have also added the posteriors of the Hubble
parameter $\Hzero$ and the cosmological constant $\OmegaL$, which are
derived from the ones we are sampling on. The slow-roll posteriors for
$\{\Pstar,\epsstar{1},\epsstar{2},\epsstar{3}\}$ also match with those
derived in~\cite{Ade:2013uln}, up to our prior which restrains
$\epsstar{3}$ to small values in order to ensure the consistency of
the slow-roll approximation. In order to make contact with the power
law spectra analysis, let us notice that $\epsstar{1}$ is only bounded
from above as it gives the tensor-to-scalar ratio $r=16
\epsstar{1}+\order{\epsilon^2}$ whereas $\epsstar{2}$ is well
constrained because it carries most of the spectral index dependency
$\nS = 1 - 2\epsstar{1} - \epsstar{2} + \order{\epsilon^2}$. On the
other hand, one sees that $\epsstar{3}$ is not constrained, and
exhibits at most a slight preference for positive values. This is also
expected since it is linked to the running of the spectral index,
which is not detected by Planck.

\subsection{Effective likelihood}
\label{sec:leff}
From equation~\eqref{eq:likeff}, the effective likelihood $\calLeff$
is simply the four-dimensional marginalised probability distribution
in the (second order) slow-roll parameter space
$\{\Pstar,\epsstar{1},\epsstar{2},\epsstar{3}\}$. It can be
straightforwardly obtained from the previous CMB slow-roll analysis,
while being more difficult to represent in a figure. At that point,
one may nevertheless question the relevance of keeping $\epsstar{3}$
in $\calLeff$ since this parameter remains unconstrained by the Planck
data. In fact, nothing prevents us to marginalise
equation~\eqref{eq:margeff} over any slow-roll parameter in addition
to the cosmological ones. Doing so over $\epsstar{3}$ implies that the
effective likelihood of equation~\eqref{eq:likeff} is now defined by
marginalisation over $\{\bthetacosmo,\epsstar{3}\}$. Let us
immediately stress that this is the correct Bayesian way to include
any uncertainties coming from the unconstrained second order terms in
the primordial power spectra, and that such an approach should become
the standard lore to perform robust inference on the first order
parameters.

From the previous discussion, we therefore consider a
three-dimensional effective likelihood obtained by marginalising over
all the $\{\bthetacosmo\}$ parameters of
equation~\eqref{eq:camparams}, plus $\epsstar{3}$, thereby ending up
with $\likeff{\bdata}{\Pstar,\epsstar{1},\epsstar{2}}$. Because this
volumic function is at the basis of all our subsequent analysis, it
has to be continuously defined for any values of its arguments. Being
only known numerically, at a set of irregular discrete points obtained
from the MCMC sampling, we have used basic machine-learning algorithms
to numerically approximate its shape. In particular, we have tested
both a radial basis function decomposition based on polyharmonic
splines~\citep{Broomhead:1988} and a more standard multivariate
interpolation using a modified quadratic Shepard's
method~\citep{Shepard:1968, Thacker:2010}. The dimension remaining
small, both methods were found to be accurate (see below) and fast:
one evaluation of the likelihood requiring typically a few
milliseconds on a standard laptop. An obvious limitation of this
method comes from the finite number of samples obtained from the MCMC
exploration: it is genuinely impossible to interpolate the likelihood
in the regions in which it takes extremely low values. As one can
guess, this is not very important because those regions have precisely
no weight in the inference process. Typically, the lowest values for
$\ln(\calLeff^{\min}/\calLeff^{{\max}}) = -10$ were obtained with the
Shepard's interpolation and this is the one we are considering in the
following.
\begin{figure}
\begin{center}
\includegraphics[width=0.49\wsmallfig]{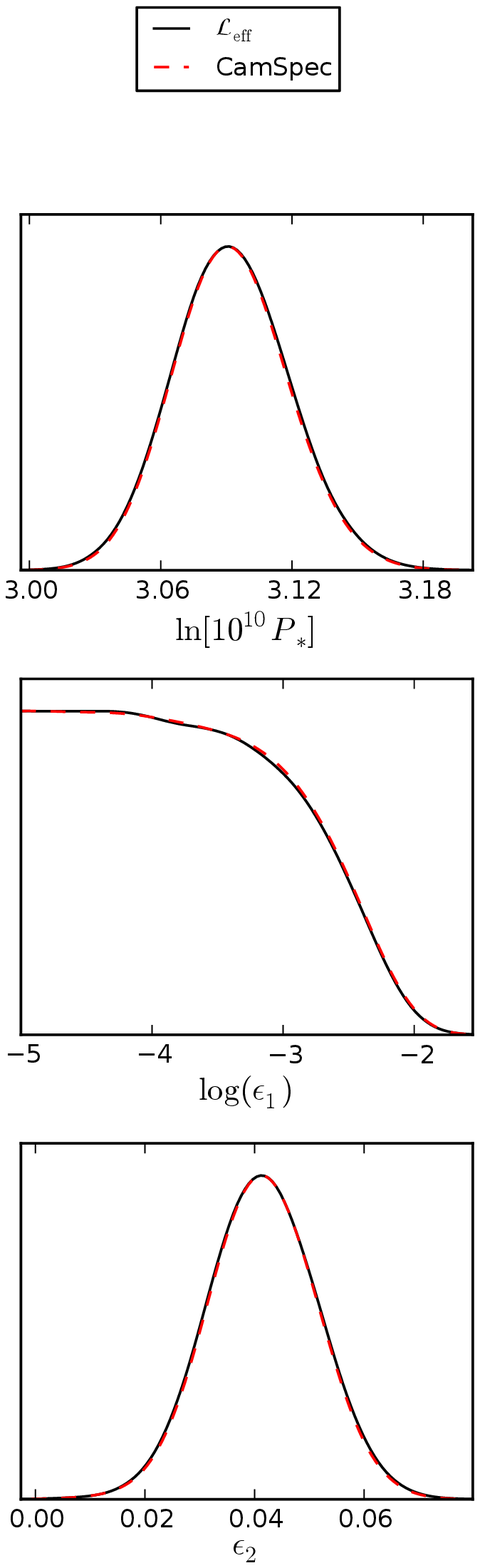}
\includegraphics[width=0.481\wsmallfig]{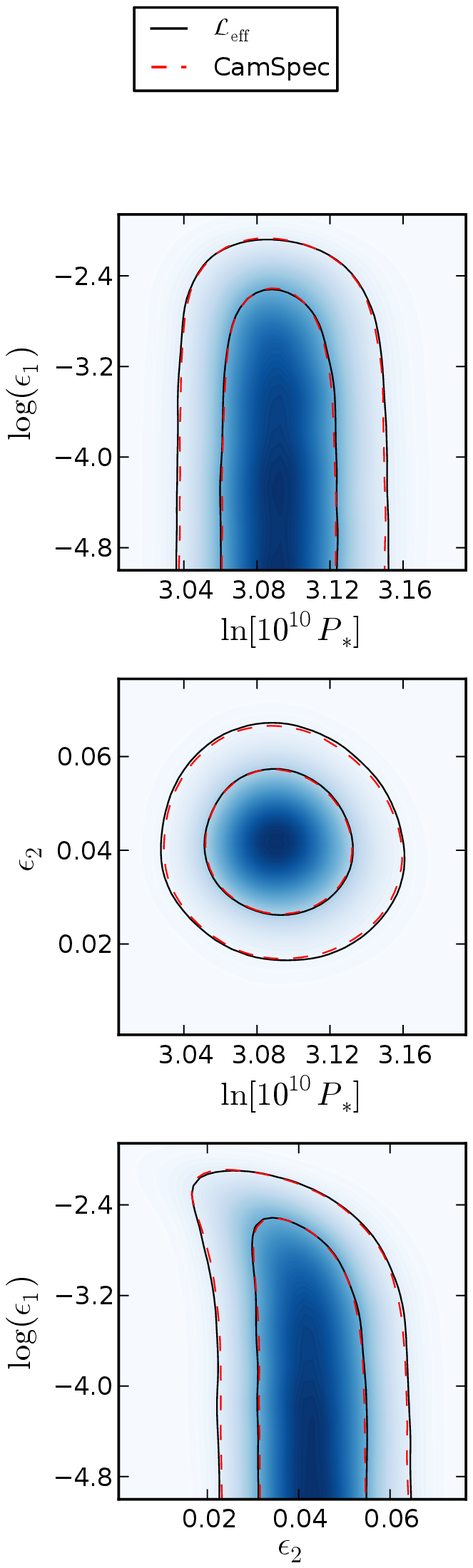}
\caption{One and two-dimensional marginalised posteriors on $\Pstar$,
  $\epsstar{1}$ and $\epsstar{2}$ obtained from a complete MCMC
  exploration using the Planck likelihood (same as in
  figure~\ref{fig:cmb}) compared to the ones coming from nested
  sampling on our effective likelihood $\calLeff$. The differences are
  barely visible.}
\label{fig:checksr}
\end{center}
\end{figure}
In order to test the accuracy of the various numerical methods
underlying the determination of $\calLeff$, we have re-derived from
scratch the marginalised posterior distributions of the amplitude
$\Pstar$ and the slow-roll parameters $\epsstar{1}$, $\epsstar{2}$
using only the machine-learned
$\likeff{\bdata}{\Pstar,\epsstar{1},\epsstar{2}}$. In order to avoid
any systematic, the slow-roll parameter space has been re-explored
using the nested sampling algorithm implemented in the publicly
available code $\MULTINEST$~\citep{Feroz:2007kg}. In order to ensure a
good convergence of the nested sampling, we have chosen the
$\MULTINEST$ convergence criteria as in~\cite{Feroz:2007kg}, i.e. a
number of live points equals to $20000$ and a target accuracy on the
global likelihood (evidence) equals to $10^{-4}$. In total, the nested
sampling of $\calLeff$ converged with a few hundred thousand samples
in ten minutes on a standard laptop. The priors for the slow-roll
parameters have been fixed as before (see section~\ref{sec:srplanck})
and we have compared in figure~\ref{fig:checksr} the one- and
two-dimensional marginalised posterior distributions coming from both
$\calLeff$ and the previous full CMB analysis. As these plots
emphasize, up to some very small deviations coming from the sampling
uncertainties, there are no differences.

\subsection{Parameter estimation for large field inflation}

In the previous section, we have shown that our numerical
implementation accurately reproduces the effective likelihood
$\calLeff$ in the slow-roll parameter space. We can now use it to
perform parameter estimation within a given inflationary scenario and,
as a proto-typical case, one can consider the ``large field
models''. These models are not currently favoured by the Planck data,
but they have the advantage that everything can be worked out
analytically thereby emphasizing the functional link between the
slow-roll parameter space and the large field one. Moreover, precisely
because they are not too close to the best fit region, these scenarios
probe a region in our likelihood $\calLeff$ which could be
problematic. In other words, they constitute a good test case for the
approach advocated here.

\subsubsection{Reheating consistent slow-roll functionals}

The large field potential is given by
\begin{equation}
V(\phi) = M^4 \left(\dfrac{\phi}{\Mpl} \right)^p,
\label{eq:pot}
\end{equation}
and, within the slow-roll approximation, the first two Hubble flow
functions read
\begin{equation}
\epsilon_1=\dfrac{p^2}{2 x^2}\,, \qquad \epsilon_2=\dfrac{2p}{x^2}\,,
\label{eq:sr}
\end{equation}
where $x \equiv \phi/\Mpl$. The field evolution is obtained by solving
the Friedmann-Lema\^{\i}tre and Klein-Gordon equations, again within
the slow-roll approximation, and one obtains
\begin{equation}
N - \Nend \simeq - \int_{\xend}^x \dfrac{V(x)}{V'(x)}
\ud x = \dfrac{1}{2p} \left(\xend^2 - x^2 \right).
\label{eq:traj}
\end{equation}
As before, the quantity $N\equiv \ln a$ is the number of {\efold} and
$\xend$ stands for the field values at which inflation ends. This equation
can be inverted to give the field value $x$ in terms of $N$ as
\begin{equation}
x = \sqrt{\xend^2 - 2p(N - \Nend)}\,.
\end{equation}
By definition, the end of inflation occurs at $\epsilon_1(\xend)=1$,
i.e. for
\begin{equation}
\xend= \dfrac{p}{\sqrt{2}}\,.
\label{eq:xend}
\end{equation}
The only quantity that remains to be determined is $\xstar$ (or
$\Nstar$), namely the field value at which the pivot mode $\kstar$
crossed the Hubble radius during inflation. Introducing the reheating
parameter~\citep{Martin:2006rs,Martin:2010kz,Martin:2010hh,Ringeval:2013hfa}
\begin{equation}
\Rrad \equiv \dfrac{\aend}{\areh}\left(\dfrac{\rhoend}{\rhoreh}
\right)^{1/4},
\end{equation}
one has
\begin{equation}
1+\zend = \dfrac{1}{\Rrad} \left(\dfrac{\rhoend}{\rhotildegamma}
\right)^{1/4}\,.
\label{eq:zend}
\end{equation}
The index ``end'' and ``reh'' denote the end of inflation and the end
of the reheating era, respectively. The reheating parameter $\Rrad$
measures any deviations the expansion of the Universe may have during
reheating compared to a pure radiation-like era. In the latter situation
$\Rrad=1$ and the reheating era cannot be distinguished from the
subsequent radiation dominated era. The quantity $\rhotildegamma
\equiv \rdofreh \rhogamma$ is the energy density of radiation today
eventually rescaled by $\rdofreh \equiv \gszero^{4/3}
\greh/(\gsreh^{4/3} \gzero)$, the change of relativistic degrees of
freedom between the reheating era and today. There, $\gs$ and $g$
respectively denote the number of entropic and energetic relativistic
degrees of freedom. By definition of the pivot scale, one has
\begin{equation}
- \kstar \etastar \simeq \dfrac{\kstar}{a(\Nstar)H(\Nstar)} =
\dfrac{\kstar}{a_0} \dfrac{a_0}{\aend}
\dfrac{\ee^{\Nend-\Nstar}}{\Hstar} = 1.
\end{equation}
From equation~\eqref{eq:zend}, making use of the
Friedmann-Lema\^{\i}tre equations, this expression can be recast into
\begin{equation}
\begin{aligned}
\Delta \Nstar \equiv \Nstar-\Nend &= -\ln\Rrad  + \dfrac{1}{4}\ln\left[
  \dfrac{9}{\epsstar{1}(3-\epsend)} \dfrac{\Vend}{\Vstar}
  \right] \\ & +\Nzero  - \dfrac{1}{4} \ln(8\pi^2 \Pstar) + \order{\epsilon},
\end{aligned}
\label{eq:efolds}
\end{equation}
in which $\Nzero \equiv \ln[(\kstar/a_0)/\rhotildegamma^{1/4}]$
roughly measures the number of {\efolds} of deceleration. We have now
at our disposal all the equations needed to determine uniquely the
observable slow-roll parameters. For any input of $\Rrad$, one can
solve the algebraic equations~\eqref{eq:efolds}, using the potential
\eqref{eq:pot} and the Hubble flow expressions \eqref{eq:sr} together
with the trajectory \eqref{eq:traj}. The solution gives $\xstar$, or
equivalently $\Delta\Nstar$, from which one gets
$\epsstar{i}(\Pstar,\Rrad,p)$, i.e. the explicit functional relation
linking the large field parameters to the slow-roll parameters.

In the previous equations, one can check that the potential parameter
$M$ cancels out. In fact, its dependency is implicit because it is in
one-to-one correspondence with $\Pstar$. This can be seen from the
first Friedmann-Lema\^{\i}tre equation, evaluated at $N=\Nstar$. In
reduced Planck units ($\Mpl=1$), defining $\vstar \equiv \Vstar/M^4 =
\xstar^p$, one has
\begin{equation}
\Hstar^2 = \dfrac{\Vstar}{3-\epsstar{1}} \simeq
M^4 \dfrac{\vstar}{3} + \order{\epsilon}.
\end{equation}
From the expression of the primordial power spectrum
\eqref{eq:powerz}, $\Hstar^2 = \Pstar (8\pi^2 \epsstar{1})
+\order{\epsilon^2}$ and one finally gets
\begin{equation}
M^4 = 24 \pi^2 \dfrac{\epsstar{1}}{\vstar}\Pstar  + \order{\epsilon^2}.
\label{eq:cobe}
\end{equation}
Given $\Pstar$ and $\xstar$, this expression completely fixes the
parameter $M$. In fact, it is more convenient to sample the parameter
space using $\Pstar$ instead of $M$ because, as argued before, it is a
well constrained quantity. In any case, one can always extract the
posterior of $M$ from the one of $\Pstar$ by using
equation~\eqref{eq:cobe}.

Concerning the reheating, different choices are possible. One can
sample directly on to the parameter $\Rrad$ introduced before, but, as
can be seen in equation~\eqref{eq:efolds}, $\Rrad$ exhibits some
explicit dependence in $\Pstar$ which will induce unnecessary
correlations in the parameter space. Following~\cite{Martin:2006rs},
it is more convenient to sample on the rescaled reheating parameter
$\Rreh$ defined by
\begin{equation}
\Rreh \equiv \Rrad \dfrac{\rhoend^{1/4}}{\Mpl}\,.
\end{equation}
Plugging this expression into equation~\eqref{eq:efolds} yields, after
some algebra with the Friedmann-Lema\^{\i}tre equations, a very
similar expression~\citep{Martin:2010kz}
\begin{equation}
\Delta \Nstar = -\ln\Rreh + \dfrac{1}{2} \ln
\left[\dfrac{9}{3-\epsend} \dfrac{\Vend}{\Vstar}\right] + \Nzero +
\order{\epsilon},
\label{eq:Refolds}
\end{equation}
which does no longer depends on $\Pstar$ (and with $\epsend=1$ here).

From the previous considerations, we will define the large field
inflationary parameter space to be
\begin{equation}
\bthetaprim=\left\{\Pstar, \Rreh, p\right\}.
\label{eq:primlfi}
\end{equation}

\subsubsection{Posteriors on the large field parameters from $\calLeff$}
\label{sec:posteff}
We have used $\MULTINEST$ to sample the parameter space of the large
field models using the numerically approximated $\calLeff$ discussed
in section~\ref{sec:leff}, together with the reheating consistent
slow-roll expressions derived in the previous section. The convergence
criteria have been chosen as before, namely $20000$ live points and a
target accuracy on the evidence equals to
$10^{-4}$~\citep{Feroz:2007kg}.  The priors have been chosen flat on
the power index $p\in[0.2,5]$, with a Jeffreys' prior on
$\ln\left(10^{10} \Pstar \right) \in [2.7,3.4]$ and another Jeffreys'
prior on the rescaled reheating parameter $\ln\Rreh\in[-46,15]$. This
last choice comes from the theoretical requirements of having the mean
equation-of-state parameter during reheating $-1/3<\wrehbar<1$, and by
imposing to the reheating energy density
$\rhonuc<\rhoreh<\Mpl^4$. Here, $\rhonuc^{1/4}\equiv 10\,\MeV$ stands
for the Big-Bang Nucleosynthesis energy scale. In addition to these
priors, we have added a ``hard prior'' which enforces that $\rhoreh
\le \rhoend$ since the reheating era necessarily starts after the end
of inflation. Those priors are discussed in more details
in~\cite{Martin:2010kz,Martin:2010hh}. In order to solve the algebraic
equation~\eqref{eq:Refolds}, we have used the public code $\ASPIC$
which also computes the needed values of
$\epsstar{i}(\Pstar,\Rreh,p)$.
\begin{figure}
\begin{center}
\includegraphics[width=\wsmallfig]{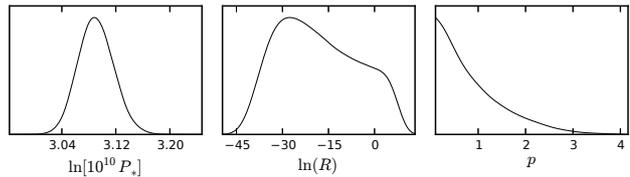}
\caption{Marginalised posterior probability distributions obtained by
  nested sampling on $\calLeff$ using the reheating consistent
  slow-roll functionals $\epsstar{i}(\Pstar,\Rreh,p)$ for large field
  inflation ($\ASPIC$). One finds the power index $p<2.3$ and the
  reheating parameter $-37 <\ln \Rreh<6$, both at $95\%$ of confidence.}
\label{fig:lfi}
\end{center}
\end{figure}
Convergence have been achieved with a few hundred thousand samples,
and in about thirty minutes on a standard laptop. The one- and
two-marginalised posteriors obtained from $\calLeff$ have been plotted
in figure~\ref{fig:lfi}. We recover that these models are under
pressure, the power index $p$ being constrained to be small. At $95\%$
of confidence, we have $p < 2.3$ and $-37 <\ln \Rreh < 6$. These
results suggest that, \emph{marginalised over all possible reheating
  history}, massive inflation is still compatible with the Planck
2013 data.

\subsection{Comparison with exact methods}

\begin{figure}
\begin{center}
\includegraphics[width=\wsmallfig]{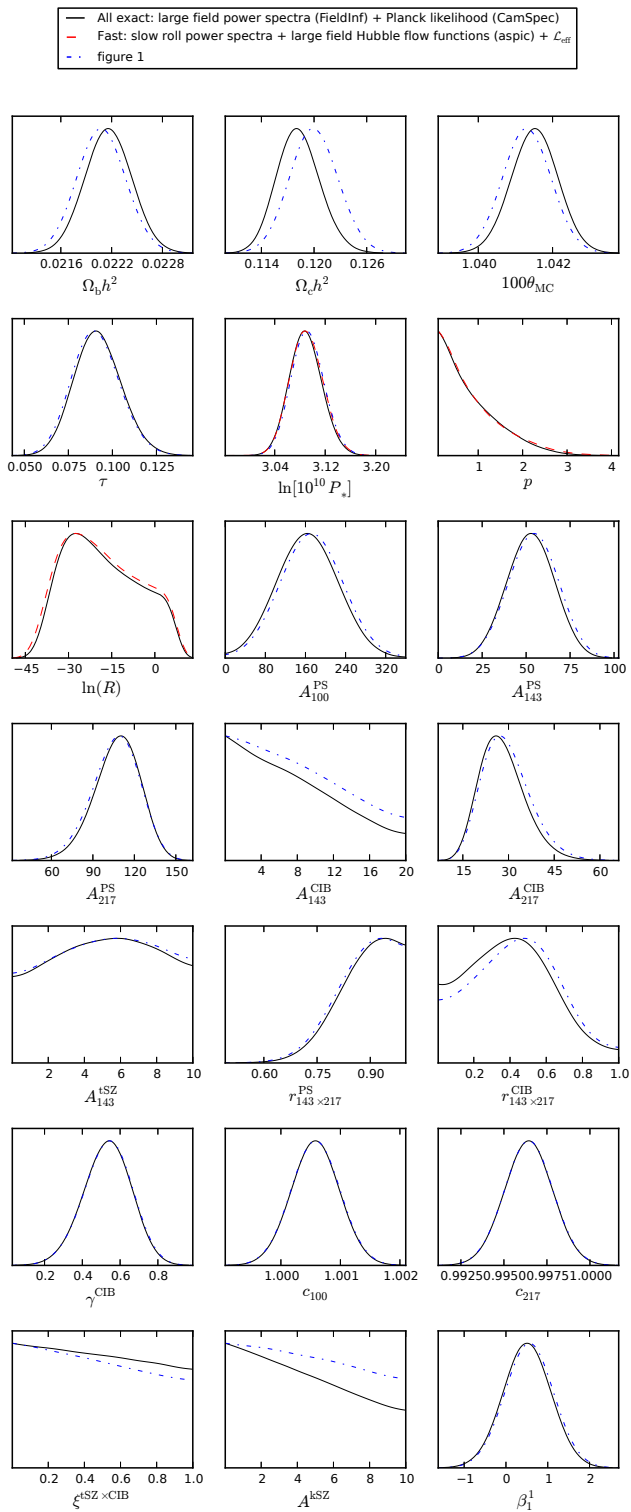}
\caption{Marginalised posterior probability distributions for all the
  parameters associated with large field inflation. The solid black
  lines are obtained from the full Planck likelihood and the exact
  primordial power spectra obtained by a mode-by-mode integration
  using the $\FIELDINF$ code. The posteriors of figure~\ref{fig:lfi},
  stemming from the effective likelihood and approximate slow-roll
  relations, have been reported as red dashed lines for $\Pstar$,
  $\ln(\Rreh)$ and $p$. The agreement is excellent. For completeness,
  we have also reported the posteriors of the cosmological parameters
  found in figure~\ref{fig:cmb} as blue dot--dashed lines. As expected,
  some of them differ due to the different primordial priors used (see
  text).}
\label{fig:largef}
\end{center}
\end{figure}

The ultimate validation of the method is to compare the previous
results to a CMB data analysis based on the exact Planck likelihood
plus the exact primordial power spectra obtained by a mode-by-mode
numerical integration of the perturbations during large field
inflation. For this purpose, we have used the public code
$\FIELDINF$\footnote{\url{http://cp3.irmp.ucl.ac.be/~ringeval/fieldinf.html}}
to integrate the scalar and tensor perturbations during inflation. The
exact power spectra are then used within a modified version of the
$\CAMB$ code which, coupled to $\COSMOMC$ and the $\CAMSPEC$
likelihood, allows us to perform a complete CMB data analysis within
large field inflation.

For the primordial parameters $\Pstar$, $\ln\Rreh$ and $p$, the priors
have been fixed exactly as in section~\ref{sec:posteff} while the
priors of all the other cosmological parameters have been chosen as in
section~\ref{sec:srplanck}. Such an analysis is in all point identical
to the one performed by~\cite{Martin:2010kz} using the WMAP seven
years data~\citep{Larson:2010gs,Jarosik:2010iu}. In particular, it is
numerically quite demanding. Requiring the $R$-statistics convergence
to be smaller than $5 \times 10^{-3}$ has taken a few thousand hours
of CPU-time on current x86-64 machines. In total, the posteriors
plotted in figure~\ref{fig:largef} have been obtained from $500000$
MCMC samples. In this figure, we have superimposed the posteriors of
$\Pstar$, $\ln \Rreh$ and $p$ coming from the fast analysis based on
slow-roll and $\calLeff$ (see figure~\ref{fig:lfi}). There is no
difference, at least for $\Pstar$ and the power index $p$. Only a
hardly visible systematic up shift on the posterior of $\ln\Rreh$
seems to be present. If not due to differences between nested sampling
and MCMC used for the fast and the exact method, respectively, it may
come from some inaccuracies of the slow-roll approximation to
determine $\xend$. Indeed, because solving $\epsilon_1(\xend)=1$ is
manifestly violating the slow-roll approximation, it is well known
that equation~\eqref{eq:xend} may induce some systematic errors of a
few {\efolds} compared to the exact field trajectory. As a result,
this translates into a systematic shift of the slow-roll approximated
value of $\rhoend$, and as such, could affect $\ln \Rreh$. This could
be easily improved, for instance by solving numerically the field
value of $\xend$, although the bias induced on the posterior is, by
far, of negligible importance with the current data. Let us notice
that this concerns only models in which inflation ends at $\epsend=1$,
and not the model for which the reheating is triggered by a tachyonic
instability ($\epsend \ll 1$ for those).

Marginalising over everything, we find the exact integration to give
almost exactly the same two-sigma confidence intervals for the
inflationary parameters. One gets $p<2.2$ (instead of $p<2.3$) and
$-37 <\ln\Rreh<6$ (unchanged) at $95\%$ confidence. Interestingly
enough, the upper limit on $p$ is not better than the one coming from
the WMAP data in~\cite{Martin:2010kz}. This may seem surprising at
first, but, in the more usual language of power law power spectra,
this can be traced back to a small shift in the best fit value of the
spectral index $\nS$ from WMAP to PLANCK. Notice that, on the
contrary, the reheating parameter $\Rreh$ is bounded from above by the
Planck data while it was only limited from below by WMAP.

Let us finally stress that our method is dedicated to perform
inference in the parameter space of inflation. As such, information on
the standard cosmological parameters has been lost into the
marginalisation process, but this is by choice because we were not
interested in this issue here. For completeness, in
figure~\ref{fig:largef}, we have superimposed the posteriors of
figure~\ref{fig:cmb} for the $\{\bthetacosmo\}$ parameters. The only
difference between these posteriors come from the primordial power
spectra: either they are second order slow-roll, or exactly integrated
within large field inflation. Precisely because the large field models
generically induce a too large tensor-to-scalar ratio, they badly fit
the data and the most probable values for some of the cosmological
parameters are accordingly shifted. Those effects cannot be studied
within the above-described effective likelihood approach. One may
nevertheless imagine to move one parameter, or more, from the set
$\{\bthetacosmo\}$ to the set $\{\bthetaprim\}$ in order to keep trace
of them in the effective likelihood.

We conclude that using our effective likelihood for inflation,
together with the slow-roll approximation, is accurate enough to deal
with data sets as precise as those from the Planck satellite.

\section{Conclusion}
\label{sec:conclusion}

In this paper, we have presented a new method to perform inference on
some primordial parameters $\{\bthetaprim\}$ associated with any
model of slow-roll inflation. It relies on the determination of an
effective likelihood $\calLeff$ for inflation which is obtained by
marginalisation over all the other parameters. This is summarized by
equations~\eqref{eq:margeff} and \eqref{eq:likeff}. The effective
likelihood being generically of much lower dimension, we have shown
that it could be easily ``machine-learned'' thereby allowing for its
fast numerical evaluation. Moreover, the low dimension of the problem
also accelerates any Bayesian exploration of the primordial parameter
space such that, in total, the speed-up reaches a few orders of
magnitude compared to an exact method.

Here, we have used the slow-roll approximation to compress information
on the primordial power spectra into a small set of parameters made of
the primordial amplitude $\Pstar$ and the Hubble flow functions
$\epsstar{i}$. For this reason, our approach can only be applied to
the slow-rolling models of inflation precisely because they are
accurately described by this approximation. As we have argued, even
within this framework, the effective likelihood space could be
extended to include any additional parameters we would like to perform
inference on, eventually including some of the cosmological ones.

Let us also mention that integrating $\calLeff$ over the parameters
$\bthetaprim$ gives the global likelihood of any slow-roll models. In
particular, this opens the window of fastly extracting and comparing
the statistical evidences of all slow-roll inflationary
models~\citep{Trotta:2005ar}.

One may be worried that an effective likelihood for inflation can only
be derived within slow-roll inflation. This is not the case. As long
as one is able to determine a set of parameters modelling accurately
the shape of the primordial power spectra, one can extract an
effective likelihood over those parameters. In fact, we could readily
extend our method to any model of inflation by binning the primordial
power spectra over a given set of modes $\{\bk\}$, and use some
efficient machine-learning algorithms to fit its multidimensional
shape, such as the recently released code $\SKYNET$
by~\cite{Graff:2011gv, Graff:2013cla}. From this, one could perform
Bayesian inference on any inflationary models by using an exact
integration code, such as $\FIELDINF$, to evaluate the power spectra
bin per bin. Certainly this would not be as fast as using the
slow-roll approximation, but by limiting the number of bins, the
speed-up would still be significant.

Finally, although we have used the Planck data as a motivated test
case, any cosmological data sets, or union of them, can be used in the
definition of $\calLeff$. Equally, the approach is not limited to the
power spectra and can be extended to other primordial observables,
such as the bispectrum and trispectrum.

\section*{Acknowledgments}
It is a pleasure to thank J.~Martin and V.~Vennin for enlightening
discussions. I would also like to thank the organisers and
participants of the CosmoStats2013 conference where some ideas
advocated here find their roots. This work is supported by the ESA
Belgian Federal PRODEX Grant No.~4000103071 and the Wallonia-Brussels
Federation Grant ARC No.~11/15-040.

\bibliography{references}

\label{lastpage}
\end{document}